\documentstyle[prd,aps,amsfonts]{revtex}

\let\NoAbstract0

\def\CC{{\Bbb C}}

\def\OO{{\Bbb O}}

\def\tr{{\rm tr}}
\def\tilde{\widetilde}
\def\bar{\overline}
\def\SpinHalf{\hbox{spin-${1\over2}$}}

\begin{document}


\title{
{\hfill\normalsize Submitted to Mod.\ Phys.\ Lett.\ {\bf A}.}\\[1.5mm]
{\bf DIMENSIONAL REDUCTION
\footnote{\ifpreprintsty\xpt\fi This essay received Honorable Mention in the
1998 competition of the Gravity Research Foundation.}
}
}

\author{Corinne A. Manogue
\footnote{\ifpreprintsty\xpt\fi
Electronic address: {\tt corinne{\rm @}physics.orst.edu}}
}
\address{
Department of Physics, Oregon State University,
		Corvallis, OR  97331, USA \\[-1.5mm]
}

\author{Tevian Dray
\footnote{\ifpreprintsty\xpt\fi
Electronic address: {\tt tevian{\rm @}math.orst.edu}}
}
\address{
Department of Mathematics, Oregon State University,
		Corvallis, OR  97331, USA \\[2mm]
}

\date{March 31, 1998}

\maketitle

\if\NoAbstract0{
\begin{abstract}
Using an octonionic formalism, we introduce a new mechanism for reducing 10
spacetime dimensions to 4 without compactification.  Applying this mechanism
to the free, 10-dimensional, massless (momentum space) Dirac equation results
in a particle spectrum consisting of exactly 3 generations.  Each generation
contains 1 massive \SpinHalf\ particle with 2 spin states, 1 massless
\SpinHalf\ particle with only 1 helicity state, and their antiparticles ---
precisely one generation of leptons.  There is also a single massless
\SpinHalf\ particle/antiparticle pair with the opposite helicity and no
generation structure.
\par
We conclude with a discussion of some further consequences of this approach,
including those which could arise when using the formalism on a curved
spacetime background, as well as the implications for the nature of spacetime
itself.
\end{abstract}

\ifpreprintsty\newpage\else{\bigskip\bigskip}\fi
}\else\bigskip\fi

One of the most attractive aspects of superstring theory is its essential
uniqueness.  Compactification, the current mechanism for reducing 10 spacetime
dimensions to 4, destroys this uniqueness, as it leads to an infinite number
of possible 4-dimensional theories.  Here we propose a new dimensional
reduction mechanism, without compactification, and apply it for simplicity to
the free massless (momentum space) Dirac equation in 10 dimensions; the result
is a unique 4-dimensional theory with some realistic physical properties.

Many authors have used the octonions to describe physics in 10 dimensions;
see for example \cite{Dixon,Gursey} and references cited therein.  In this
formalism, the Dirac equation can be written
\begin{equation}
\tilde{P}\psi = 0
\label{MomDirac}
\end{equation}
where $\psi$ is a 2-component octonionic column representing a Majorana-Weyl
spinor, $P = p^\mu \sigma_\mu$ is a $2\times2$ Hermitian octonionic matrix
representing a vector, $\tilde{P} := P - \tr(P) I$, and the $\sigma_\mu$ are
generalized Pauli matrices.  In this equation, the role of the gamma matrices
is taken by octonionic multiplication.

As discussed in \cite{FairlieII,Sudbery,Thesis,Superparticle,Clifford}, $p^\mu
p_\mu \equiv -\det(P)=0$, which implies the existence of a 2-component
octonionic spinor $\theta$ such that
\begin{equation}
P = \pm\theta\theta^\dagger
\label{SolI}
\end{equation}
where the sign depends on the time orientation, i.e.\ on the sign of $p^0$.
The components of $\theta$ can be chosen to lie in the complex subspace of
$\OO$ determined by $P$.  The general solution of (\ref{MomDirac}) is then
\begin{equation}
\psi = \theta\xi
\label{SolII}
\end{equation}
where $\xi$ is an arbitrary octonion.

The $SO(9,1)$ invariance of (\ref{MomDirac}) can be seen using the explicit
representation of {\it finite} Lorentz transformations in 10 spacetime
dimensions in terms of $SL(2,\OO)$ given by Manogue \& Schray
\cite{Thesis,Lorentz}; this is is more usually discussed at the Lie algebra
level \cite{Tony,Chung}.

We now choose a preferred octonionic unit $\ell$; we label the octonionic
units $i,j,k,k\ell,j\ell,i\ell,\ell$, choosing $\ell$ rather than $i$ for the
preferred unit and saving $i$, $j$, $k$ for a distinguished quaternionic
triple.  The map
\begin{equation}
\pi(q) = {1\over2} (q + \ell q \bar\ell) .
\label{Project}
\end{equation}
projects $\OO$ to a preferred complex subalgebra $\CC\subset\OO$, thus
determining a preferred $SL(2,\CC)$ subgroup of $SL(2,\OO)$ and breaking the
10-dimensional Lorentz invariance down to 4-dimensional Lorentz invariance.
This is the heart of the new dimensional reduction mechanism.

What are the physical consequences of this dimensional reduction?

We first identify the possible particles using octonionic versions of the
standard energy and spin projections.  Since $P$ is null ($\det(P)=0$), the
projected vector field $\pi(P)$ is timelike or null ($\det(\pi(P))\ge0$), and
has the same time-orientation as $P$.  We therefore obtain both massive and
massless particles, with mass given by
\begin{equation}
\det\big(\pi(P)\big) = m^2 .
\end{equation}
We now use a (4-dimensional) Lorentz transformation to bring a massive
particle to rest, or to orient a massless particle in the $z$-direction.

We can distinguish particles from their antiparticles by the sign of $p_0$, as
usual; this is the same as the sign on the RHS of (\ref{SolI}).  For massive
particles, we can equivalently distinguish particles from antiparticles using
the energy projections
\begin{equation}
\Pi_\pm = {1\over2} \left( \sigma_t \mp \sigma_k \right)
\end{equation}
where $\sigma_t=I$ and where, without loss of generality, we have chosen our
octonionic basis so that $P-\pi(P)$ is a multiple of $\sigma_k$, the
(generalized) Pauli matrix corresponding to $k$.

Due to the lack of commutativity and associativity, the spin operator takes
the form
\begin{equation}
\hat{L}_z \psi := - (L_z \psi) \ell
\end{equation}
where
\begin{equation}
L_z = {1\over2} \pmatrix{\ell& 0\cr \noalign{\smallskip} 0& -\ell\cr} .
\end{equation}
as usual.  $\hat{L}_z$ is self-adjoint \cite{Eigen} with respect to the inner
product
\begin{equation}
\langle \psi,\chi \rangle = \pi \!\left( \psi^\dagger\chi \right)
\end{equation}
and spin eigenstates are obtained as the eigenvectors with eigenvalues
$\pm{1\over2}$.

Putting this all together, we find 1 massive \SpinHalf\ particle at rest, with
2 spin states, namely
\begin{equation}
e_{\scriptscriptstyle\uparrow} = \pmatrix{~~1\cr -k\cr}
\qquad
e_{\scriptscriptstyle\downarrow} = \pmatrix{k\cr 1\cr}
\label{Electron}
\end{equation}
whose antiparticle is obtained by replacing $k$ by $-k$ (and changing the sign
in (\ref{SolI})).  These states (and those below) may be multiplied on the
right by an arbitrary complex number.  We also find 1 massless \SpinHalf\
particle involving $k$ moving in the $z$-direction, with a single helicity
state
\begin{equation}
\nu_z = \pmatrix{0\cr k\cr}
\label{Neutrino}
\end{equation}
which represents both a particle and its antiparticle, depending on the sign
in (\ref{SolI}).  It is important to note that
\begin{equation}
\nu_{\hbox{-}z}=\pmatrix{k\cr 0\cr}
\end{equation}
corresponds to the same massless particle moving in the opposite direction but
with the same helicity!

There is also a single {\it complex} massless \SpinHalf\ particle, with the
opposite helicity, given by
\begin{equation}
\hbox{\O}_z=\pmatrix{0\cr 1\cr} .
\end{equation}
As with
the other massless states, this describes both a particle and an antiparticle.

We have shown how the massless Dirac equation in 10 dimensions reduces to the
(massive and massless) Dirac equation in 4 dimensions.  Choosing a preferred
octonionic unit $\ell$ not only reduces 10 dimensions to 4, it also singles
out 3 natural, nonoverlapping quaternionic subalgebras of $\OO$ which contain
$\ell$, namely those associated with the preferred triple $i$, $j$, $k$.  We
identify these 3 quaternionic spaces as describing 3 generations.
Furthermore, each such generation consists precisely of 1 massive \SpinHalf\
particle with 2 spin states, 1 massless \SpinHalf\ particle with only 1
helicity, and their antiparticles.  We identify this set of particles with a
generation of leptons.  This formalism thus predicts 3 generations of leptons
with the correct number of spin/helicity states!

In addition, there is also a single massless particle/antiparticle pair, which
is purely complex.  It therefore does not belong to any generation, and it has
the opposite helicity from the other massless particles.  We do not currently
have a physical interpretation for this additional particle; if this theory is
to correspond to nature, then for some reason this additional particle must
not interact much with anything else.

A more detailed description of the theory described here will be presented
elsewhere \cite{Spin}.  The theory can also be elegantly rewritten in terms of
\hbox{$3\times3$} octonionic Hermitian matrices, similar to the approach to
the superparticle presented in \cite{Thesis,Superparticle}.  This approach,
which will be presented elsewhere \cite{Ions}, demonstrates that the theory is
invariant under a much bigger group than the Lorentz group, namely the
exceptional group $E_6$.  We are attempting to use this larger group to extend
the theory so as to include interactions, especially $SU(2)\times U(1)$ for
leptons, and possibly even color and quarks.

This formalism contains essentially only 1 free parameter per generation,
namely the length scale to be associated with the octonionic units $i$, $j$,
and $k$.  Should it be possible to include color, and hence quarks, in this
description, this would lead to definite predictions, for instance, relating
the lepton masses to the masses of (free) quarks.  Furthermore, since the
natural way to describe this scaling is in terms of a metric, the
consideration of a non-flat background metric could lead to observable
phenomena.  A non-flat background could change the orientation of the
octonionic directions from one spacetime point to another, resulting in
generation mixing which would depend on the strength of the background
curvature.  Perhaps nuclear reactions in the sun run at different rates from
those on Earth.

We have worked only in momentum space, and have discussed only free particles.
Perhaps our most intriguing result is the observation that the introduction of
position space requires a preferred complex unit in the Fourier transform.
Similarly, a description of interactions based on minimal coupling again
involves a preferred complex unit.  Therefore, it does not appear to be {\it
possible} to use the formalism presented here to give a full, interacting,
10-dimensional theory in which all 10 spacetime dimensions are on an equal
footing.  We view this as a tantalizing hint that not only interactions, but
even 4-dimensional spacetime itself, may arise as a consequence of the
symmetry breaking from 10 dimensions to 4!

\medskip
\leftline{\it Acknowledgments}

It is a pleasure to thank David Griffiths, Jason Janesky, Phil Siemens, and
Pat Welch for discussions, Paul Davies for moral support, and the physics
departments at Reed College and the University of Adelaide for hospitality.
We are grateful for the vision of the participants in the first Octoshop
(G\"oteborg, 1994), namely Rafa\l \ Ab\l amowicz, Martin Cederwall, Geoffrey
Dixon, Pertti Lounesto, Ian Porteous, and Tony Smith, several of whom also
participated in the second Octoshop (Corvallis, 1997).  This work was deeply
influenced by a long-standing collaboration with Tony Sudbery, J\"org Schray,
and especially David Fairlie, who started it all \cite{FairlieI}.  This work
was partially supported by NSF Grant PHY-9208494 (CAM \& TD) and a Fulbright
Grant (TD) under the auspices of the Australian-American Educational
Foundation.

\ifpreprintsty{}\else\vspace{-0.45cm}\fi


\begin{references}
\ifpreprintsty{}\else\vspace{-1.55cm}\fi




\bibitem{Dixon}
Geoffrey M. Dixon,
{\bf Division Algebras: Octonions, Quaternions, Complex Numbers and the
	Algebraic Design of Physics},
Kluwer Academic Publishers, Boston, 1994.

\bibitem{Gursey}
Feza G\"ursey and Chia-Hsiung Tze,
{\bf On the Role of Division, Jordan, and Related Algebras in Particle
	Physics},
World Scientific, Singapore, 1996.

\bibitem{FairlieII}
David B.~Fairlie \& Corinne A.~Manogue,
Phys.\ Rev.\ {\bf D36}, 475--479 (1987).

\bibitem{Sudbery}
Corinne A.~Manogue \& Anthony Sudbery,
Phys.\ Rev.\ {\bf D40}, 4073--4077 (1989).

\bibitem{Thesis}
J\"org Schray,
{\bf Octonions and Supersymmetry},
Ph.D.\ thesis, Department of Physics, Oregon State University, 1994.

\bibitem{Superparticle}
J\"org Schray,
Class.\ Quant.\ Grav.\ {\bf 13}, 27 (1996);

\bibitem{Clifford}
J\"org Schray \& Corinne A.~Manogue,
Foundations of Physics, {\bf 26} (1996) 17--70.

\bibitem{Lorentz}
Corinne A. Manogue and J\"org Schray,
J. Math.\ Phys.\ {\bf 34}, 3746--3767 (1993).

\bibitem{Tony}
A. Sudbery,
J. Phys.\ {\bf A17}, 939 (1987).

\bibitem{Chung}
K. W. Chung \& A. Sudbery,
Phys.\ Lett.\ {\bf B198}, 161 (1987).

\bibitem{Eigen}
It is not true that octonionic self-adjoint operators have only real
eigenvalues.  In fact, the spin eigenstates given in (\ref{Electron}) and
(\ref{Neutrino}) are simultaneous eigenstates of all 3 angular momentum
operators, albeit with quaternionic eigenvalues!  However, only real
eigenvalues correspond to observables.
For further details, see:
Tevian Dray and Corinne Manogue,
{\it The Octonionic Eigenvalue Problem},
(in preparation).

\bibitem{Spin}
Corinne A. Manogue and Tevian Dray,
{\it Quaternionic Spin},
(in preparation).

\bibitem{Ions}
Corinne A. Manogue and Tevian Dray,
(in preparation).

\bibitem{FairlieI}
David B.~Fairlie \& Corinne A.~Manogue,
Phys.\ Rev.\ {\bf D34}, 1832-1834 (1986).

\end{references}
\end{document}